\documentstyle[epsfig,prd,aps]{revtex}

\tighten


\def\ptitle#1{\wideabs{\maketitle
\abstract #1 \endabstract \pacs{Pacs numbers: 03.70.+k, 98.80.Cq}}}

\begin{document}

\draft 

\preprint{NSF-ITP-98-001\\GRP-490}

\title{Cosmic Censorship: As Strong As Ever}

\author{Patrick R. Brady$^{(1)}$,  Ian G. Moss$^{(2)}$ and Robert C.
Myers$^{(3)}$} 

\address{${}^{(1)}$Theoretical Astrophysics 130-33, California
Institute of Technology, Pasadena, CA 91125\\ 
${}^{(2)}$Department of Physics, University of Newcastle Upon Tyne,
NE1 7RU U.K.\\ 
${}^{(3)}$Institute of Theoretical Physics, University of California,
Santa Barbara, CA 93117}

\ptitle{
Spacetimes which have been considered counter-examples to strong
cosmic censorship are revisited.  We demonstrate the classical
instability of the Cauchy horizon inside charged black holes embedded
in de Sitter spacetime for all values of the physical parameters.  The
relevant modes which maintain the instability, in the regime which was
previously considered stable, originate as outgoing modes near to the
black hole event horizon.  This same mechanism is also relevant for
the instability of Cauchy horizons in other proposed counter-examples
of strong cosmic censorship.
}

\narrowtext

 As demonstrated by the elegant theorems of Hawking and
Penrose\cite{hawk}, spacetime singularities are ubiquitous features
of general relativity.  Thus Einstein's theory itself impels us to
search for a more fundamental theory of gravity in order to understand
the physics of these extreme situations. The utility of general
relativity in describing gravitational phenomena is maintained by
cosmic censorship\cite{pen}. The latter is based on the
common wisdom that singularities are not pervasive, and has been
expressed in two forms:
\begin{list}{}{\leftmargin=1.2em\itemsep=2pt\topsep=0em\parsep=0em}
\item[1.] {\it Weak Cosmic Censorship} states that, beginning
with generic initial conditions, singularities only form
in gravitational collapse hidden behind an event horizon.
\item[2.] {\it Strong Cosmic Censorship} states that the evolution
of generic initial data will always produce a globally hyperbolic
spacetime.
\end{list}

Thus the weak form of the conjecture suggests singularities are always
hidden inside of black holes, invisible to distant observers. The
strong form indicates that singularities only appear on spacelike or
null surfaces, and so are hidden from all observers, {\it i.e.,} the only way
to examine a spacetime singularity is to run into it. At present, no
rigorous theorems have been established to prove either of these
conjectures, rather the evidence for (or against) cosmic censorship
comes from our experience in solving Einstein's equations. Of the two
conjectures, weak cosmic censorship enjoys a better bill of
health\cite{wally}. Strong cosmic censorship seems to have run afoul
of certain counter-examples in which timelike singularities develop
for a (small but) finite range of physical parameters\cite{cham}.
Strong cosmic censorship and these examples are the focus of this
letter. We will demonstrate that a more complete analysis of the
latter solutions shows that they do not provide counter-examples to
strong cosmic censorship.  It is worth emphasizing that the failure of
the strong form of cosmic censorship would indicate that the
predictability of the Einstein equations can be lost in
regions of spacetime where observers encounter no extreme
gravitational fields.

Solutions of Einstein's equations which have timelike singularities
hidden inside event horizons are familiar; both 
Reissner-Nordstr\"om and Kerr-Newman black holes belong to this class.
In general, there is a Cauchy horizon (CH) associated with a timelike
singularity. The CH is a null hypersurface which marks the limit of
the evolution of the solution from some initial time slice; that is,
observers that cross the CH enter a region in which past directed null
geodesics may terminate on the singularity.  The Reissner-Nordstr\"om
solution, given by setting $\Lambda= 0$ in Eqs.~(\ref{metric}) and
(\ref{fun}), is the archetypical example of this situation.  The
solution has two horizons at $r_\pm = M\pm \sqrt{M^2-Q^2}$ determined
by solving $f(r)=0$.  The smaller horizon $r_-$ is the CH.  The
solution can be analytically extended to include $r=0$, which is then
the locus of a timelike singularity. However, many extensions to
$r<r_-$ are possible corresponding to alternative boundary conditions
at the origin. Thus this elementary solution of Einstein equations is
not globally hyperbolic.

Nevertheless, the Reissner-Nordstr\"om metric should not be considered
a counter-example to strong cosmic censorship.  Building on the initial
observation by Penrose~\cite{Penrose_R:1969} that the CH is a surface
of infinite gravitational blueshift, it has been demonstrated that the
CH is unstable to linear gravitational and electromagnetic
perturbations~\cite{Matzner_R:1979,Chandra_S:1982}.  Further
investigations have demonstrated that the CH is transformed into a
null, scalar curvature singularity when full non-linear evolution is
considered~\cite{Poisson_E:1990,Ori_A:1991,Brady_P:1995a}.  The
essential feature responsible for the instability is the same in
all of these analyses: small time-dependent perturbations
originating outside the black hole are gravitationally blueshifted as
they propagate inwards parallel to the CH.  The locally measured
flux of these perturbations grows without bound as the CH is
approached along timelike geodesics.

This situation changes if the charged black hole is immersed in de
Sitter space by the introduction of a positive cosmological constant,
$\Lambda$. The metric takes the form 
\begin{eqnarray}
	ds^2 &=& -f(r) dt^2+ {dr^2\over f(r)}
	+ r^2 (d\theta^2 +\sin^2\theta d\phi^2) \label{metric} \; ,\\
	f(r) &=& 1 - \frac{2M}{r} + \frac{Q^2}{r^2} + 
	\frac{\Lambda r^2}{3} \label{fun} \; .
\end{eqnarray}
In this solution, there are three horizons corresponding to
the positive solutions of $f(r)=0$; we label them $r_3\leq r_2\leq
r_1$ where $r_3$ denotes the Cauchy horizon, $r_2$, the event horizon,
and $r_1$ is the radius of the cosmological horizon.  Thus one again
finds an inner CH and a timelike singularity at $r=0$.  In terms of
global structure, the main modification is at large radius where the
spacetime is asymptotically de~Sitter rather than flat.  As a result,
the standard blueshift argument of Penrose is slightly modified.
Radially infalling radiation which propagates along the CH originates
in the asymptotic region close to the cosmological horizon.
Consequently, such radiation is redshifted as it falls away from the
cosmological horizon as well as being blueshifted at the CH; there is
a competition of these two effects in determining the corresponding
flux of radiation at the CH. For a limited range of physical
parameters (corresponding to near-extremal black holes), one finds
that the cosmological redshift dominates and a finite flux is
produced\cite{beep,Mellor_F:1992}.  Thus this mechanism is ineffective
in destabilizing the CH.

The essential point of the present letter is that, in this latter
situation, one
must extend the analysis to also consider outgoing modes which originate
from close to the event horizon.
These modes are scattered by the curvature to produce an additional
influx along the CH.  There is again a competition of a redshift
in climbing away from the event horizon and a blueshift in falling
towards the CH, but in this case, the latter {\it always} dominates to
produce a diverging flux at the CH. Generically, this effect is
subdominant in comparison to the flux due to the infalling modes,
however, it persists into the regime where the latter only yield a
finite flux.  This argument, which is made precise below, demonstrates
that the CH remains unstable over the entire range of physical
parameters, and that Reissner-Nordstr\"om-de~Sitter black holes are
not counter-examples to strong cosmic censorship. 

\begin{figure}  %%BoundingBox: 114 186 499 606
\psfig{file=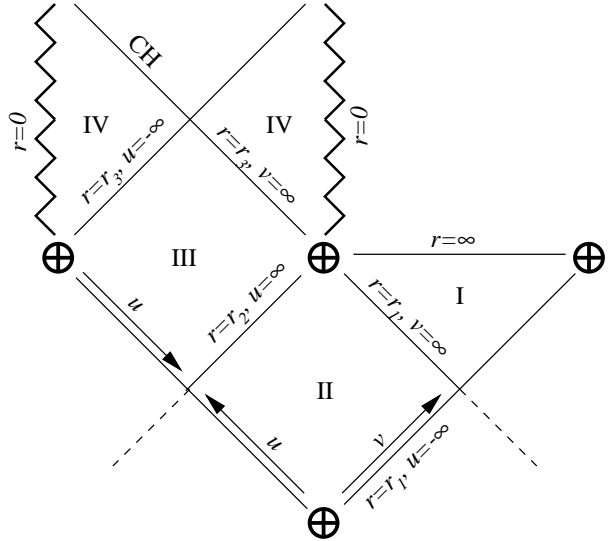,width=8cm,bbllx=114pt,bblly=186pt,bburx=544pt,bbury=606pt}
\caption{\label{fig:penrose} A portion of the Penrose conformal diagram
for the Reissner-Nordstr\"om-de~Sitter black hole spacetime.
Regions II and III correspond to the exterior and interior of
the black hole, respectively, separated by the event horizon at $r=r_2$.
See Ref.~[4]
%[cham] {\bf **fix this**}
for a detailed description of the spacetime geometry.}
\end{figure}

To begin a quantitative discussion, we transform the metric (\ref{metric})
to null coordinates
\begin{equation}
	ds^2 = -f(r) dv\, du + r^2 (d\theta^2 +\sin^2\theta d\phi^2)
\label{meaty}
\end{equation}
where $u=t-r_*$ and $v=t+r_*$ are defined in terms of the tortoise
radial coordinate
\begin{equation}
	r_*= \int dr/f(r) \; .
\label{tort}
\end{equation} 
These coordinates are illustrated in Fig.~\ref{fig:penrose}.  The main
points to note are: $v=\infty$ on the ingoing sheets of the
cosmological and the inner horizons, and $u=\infty$ on the outgoing
sheet of the black hole event horizon.  The various blueshift and
redshift effects discussed above are controlled by the surface
gravities of the respective horizons. The latter are defined
by
\begin{equation}
	\kappa_i = \frac{1}{2} \left| \frac{df}{dr} \right|_{r=r_i}
\label{surf}
\end{equation}
where $1\leq i \leq 3$. 

In this spacetime, we consider the evolution of linearized
perturbations denoted as $\Phi$.  The field $\Phi$ satisfies a wave
equation, which can be reduced to a one-dimensional scattering
problem, {\it e.g.,} see Eq.~(\ref{eq:scattering-equation}) below, by virtue
of the spherical symmetry and static nature of the background
spacetime. If the evolution produces a diverging flux of radiation as
measured by observers at the CH, the result is interpreted as
indicating the CH is unstable.  The flux received by any observer is
proportional to the square of the amplitude
\begin{equation}
	{\cal F} = \Phi_{,\alpha}\, u^\alpha \label{eq:flux}
\end{equation}
where $u^\alpha$ is the observer's four-velocity.

Now the essential features of the linear perturbation analysis can be
summarized by the following argument.  First,
reasonable initial conditions must be determined for perturbations in
the vicinity of the cosmological and event horizons.  Generally,
observers crossing the cosmological horizon will measure a finite
flux. Considering Eq.~(\ref{eq:flux}) for a radially moving observer
one shows that $\Phi$ must satisfy
\begin{equation}
	\Phi_{,v} \sim e^{-\kappa_1 v} \label{eq:initial-ingoing}
\end{equation}
as $v\rightarrow \infty$.  This determines the behavior of the initial
ingoing modes.  Observers falling into the black hole should see a
finite flux of radiation at the event horizon. Similarly, this
requires that the variation of the field satisfy
\begin{equation}
	\Phi_{,u} \sim e^{- \kappa_2 u} \label{eq:initial-outgoing}
\end{equation}
as $u\rightarrow \infty$ in {\it both} regions II and III, fixing the
initial conditions for the outgoing modes. The evolution of these
outgoing modes will result in backscattering adding an extra
contribution to the influx along the CH.  This additional flux may be
estimated by observing that the backscattering occurs roughly on a
$v-u=${\it constant} surface, {\it i.e.,} the effective potential falls off
very rapidly near the event and Cauchy horizons,
and so in this process the $u$-dependence of
Eq.~(\ref{eq:initial-outgoing}) is converted to a $v$-dependence.
Therefore, the total amplitude measured by observers crossing the CH
takes the form
\begin{equation}
	{\cal F} \sim e^{\kappa_3 v} \left( 
	e^{-\kappa_1 v} + {\rm constant} \times e^{-\kappa_2 v}
	\right) \; .
\label{amper}
\end{equation}
The first term above, due to the ingoing modes, produces a divergent
result for $\kappa_3>\kappa_1$, which is satisfied {\it except} for
near-extremal black holes\cite{beep,Mellor_F:1992}. The backscattered
contribution diverges for $\kappa_3>\kappa_2$, which is valid whenever
$r_3\neq r_2$. Therefore the second flux ensures that the CH is
generally unstable. It should be noted that over
most of the range of physical parameters $\kappa_1 >\kappa_2$, and so
the backscattered term is subdominant and neglecting the outgoing
modes still yields quantitatively correct results. It is only in the
regime previously thought to be stable, {\it i.e.,} $r_3\simeq r_2$, that the
importance of the outgoing modes manifests itself.

While the previous argument may appear simplistic, the final result
for the amplitude (\ref{amper}) is supported by our detailed analysis
of the linear instability of the Cauchy horizon. Our approach was
three-fold: extending the null fluid model of \cite{beep} and the mode
analysis of \cite{Mellor_F:1992} to incorporate backscattering, and
making numerical investigations to confirm the latter analytic
results. The details of this work will be presented elsewhere, but
here we discuss the new result revealed by the mode
analysis. This mechanism for the instability of the CH arises
purely from modes confined to the {\it interior} of the black hole,
{\it i.e.,} region III of Fig.~\ref{fig:penrose}.

The equations governing the metric and electromagnetic perturbations
of a Reissner-Nordstr\"om-de~Sitter black hole have been worked out in
detail in~\cite{Mellor_F:1990}, where it was shown that they reduce to
four scalar wave equations. The perturbation fields $\Phi$ are decomposed into 
eigenmodes of frequency $\omega$ and spherical harmonics, which satisfy
\begin{equation}
\left( \frac{d^2}{dr_*^2} + \omega^2 \right) \hat\Phi(\omega,r_*) =
V(r_*) \hat\Phi(\omega,r_*) \; . \label{eq:scattering-equation}
\end{equation}
(where angular eigenvalues will be suppressed throughout).
The details of the potential depend on the type of perturbation
\cite{Mellor_F:1990}, {\it e.g.,} for axial perturbations
\begin{equation}
V = f(r) \left[ \frac{a}{r^2} + \frac{b}{r^3} + \frac{c}{r^4}
	\right]	\; ,
\end{equation}
where Eq.~(\ref{fun}) gives $f(r)$, Eq.~(\ref{tort}) determines $r(r_*)$,
and $a,b,c$ are certain constants. An important general feature is that
the potential is always analytic in both $\exp(-\kappa_3 r_*)$ and
$\exp(\kappa_2 r_*)$ throughout region III.
It is useful to introduce a basis of mode solutions of 
Eq.~(\ref{eq:scattering-equation}): $\overleftarrow\Phi(\omega, r_*)$
and $\overrightarrow\Phi(\omega, r_*)$ 
normalized to satisfy 
\begin{eqnarray}
	\begin{array}{l}
	\overleftarrow\Phi(\omega, r_*) \rightarrow e^{-i\omega r_*}\\
	\overrightarrow\Phi(\omega, r_*) \rightarrow e^{i\omega r_*}
	\end{array} 
	\mbox{\hspace{0.2in} as } r_*\rightarrow -\infty \; . 
	\label{eq:asymptotic}
\end{eqnarray}
These modes represent initially ingoing and outgoing waves, respectively,
in the black hole interior.
The full time-dependent solution can now
be written as
\begin{eqnarray}
\Phi(t,r_*)&=& \int_{-\infty}^\infty\frac{d\omega}{2\pi}
	\left[\overleftarrow W(\omega) \overleftarrow \Phi(\omega,r_*)
\right.\nonumber\\
&&\qquad\qquad\left.+\overrightarrow W(\omega)\overrightarrow\Phi(\omega,r_*)
	\right] e^{-i\omega t} \label{field} \; ,
\end{eqnarray}
with
the functions $W(\omega)$ being determined by the initial data.

Perturbations falling in across the event horizon from the exterior
would fix $\overleftarrow W(\omega)$. These would be analyzed as in
Ref.~\cite{Mellor_F:1990}, and we do not consider them further here.
Instead we focus on outgoing perturbations which would arise from the
surface to the star which collapses to form the black hole. These
would be the perturbations determining $\overrightarrow W(\omega)$.
The asymptotic behavior of the field given in
Eq.~(\ref{eq:initial-outgoing}) implies that $\overrightarrow{W}(\omega)$ has a
pole at $\omega=-i\kappa_2$. 

As above, we wish to determine the flux of radiation measured by an
observer crossing the CH, and so must calculate the amplitude ${\cal F}$
defined in Eq.~(\ref{eq:flux}). The part of the
amplitude which may be divergent at the CH is
\begin{equation}
	{\cal F} \sim e^{\kappa_3 v}\Phi_{,v} \; .
\end{equation}
Now the initially outgoing modes of Eq.~(\ref{eq:asymptotic}) are
dispersed by the
potential between the two horizons so that as $r_*\rightarrow\infty$
\begin{equation}
	\overrightarrow{\Phi} \rightarrow {A(\omega)} e^{i\omega r_*} + 
	{B(\omega)} e^{-i\omega r_*} \; .
\end{equation} 
It is the behavior of the reflected waves that are relevant to our
discussion, thus
\begin{equation}
	\overrightarrow{\cal F} \sim e^{\kappa_3v} \int_{-\infty}^{\infty}
	\!\!\!\!d\omega \; \omega\,\overrightarrow{W}(\omega)\, B(\omega) e^{-i\omega v} \;
.
\end{equation}
The integral is computed by closing the contour in the lower
half-plane and using the residue theorem.  The dominant contribution
to the flux comes from the pole nearest to the real axis.  Using
arguments similar to those in Ref.~\cite{Chandra_S:1982}, one shows
that $\omega\, B(\omega)$ is analytic in the strip
$[-i\kappa_3,i\kappa_2]$.  Hence the pole in $\overrightarrow{W}(\omega)$ at
$-i\kappa_2$ provides the leading contribution, that is
\begin{equation}
	\overrightarrow{\cal F} \sim e^{(\kappa_3-\kappa_2)v} \left\{
	-i\kappa_2\, B(-i\kappa_2)\, {\rm Res}[\overrightarrow{W}(-i\kappa_2)]\right\} \; .
\end{equation}
As discussed above, it is easy to show that
$\kappa_3>\kappa_2$ provided that $r_3\neq r_2$, therefore, $\overrightarrow{\cal
F}$ always diverges as $v\rightarrow
\infty$ provided $B(-i\kappa_2)$ is non-zero.  While it seems unlikely
that $B(-i\kappa_2)$ would vanish, we have verified it numerically.

The evolution of scalar waves on the spherical black-hole de Sitter
spacetimes has been considered in~\cite{Brady_P:1997}.  It is
straightforward to apply the same numerical techniques to the fields $\Phi$
above, except inside the black hole horizon. Reinstating the time-dependence
in the scattering equations (\ref{eq:scattering-equation}), the wave equations
may be written as
\begin{equation}
	\Phi_{,uv} = -\frac{1}{4} V(r) \Phi \; . \label{eq:wave-equation}
\end{equation}
We use a characteristic evolution scheme to solve these equations, so
the initial data is supplied on an ingoing null hypersurface $v=0$,
and the event horizon of the black hole (in reality a very large
positive value of $u$).  The initial data corresponds to what can be
reasonably expected from a collapsing star.  Near to the event horizon
the field is presumed to be analytic in a Kruskalized coordinate
tailored to that horizon. Thus
\begin{equation}
\Phi(u,v=0) \simeq \Phi_0 + \Phi_1 e^{-\kappa_2 u} + \ldots
\end{equation}
which reproduces the dependence of Eq.~(\ref{eq:initial-outgoing}).
On the event horizon the field was taken to decay exponentially with
advanced time;  the precise form was motivated by considerations of
tails of gravitational collapse in the external field of the black
hole~\cite{Brady_P:1997}. However the results are insensitive to
the details of these boundary conditions.

The results of the numerical integration are consistent with the
scattering analysis described above. We find the rate of decay of
the field satisfies
\begin{equation}
\Phi_{,v} \propto e^{-\sigma v} 
\end{equation}
along surfaces of constant $u$ crossing the CH, where the decay constant
$\sigma$ was found to equal $\kappa_2$ within numerical errors. 
For example, in a calculation with $M=1.0$,
$Q=1.000015$ and $\Lambda=10^{-4}$, 
$\sigma$ was equal $\kappa_2$ to an accuracy of $\sim
0.035\%$.  Moreover, the decay of the perturbations at the horizon was
also found to be independent of the angular eigenvalues, in
contrast to the results obtained for wave evolution in the exterior
region \cite{Brady_P:1997}. Together these calculations show that the
instability of the CH will in fact generally result from modes
entirely confined to the interior of the black hole.

In this letter, we have shown that the CH of the
Reissner-Nordstr\"om-de~Sitter black hole is unstable to linear
perturbations over the entire range of physical parameters. This
should imply that CH instability also arises in the full nonlinear
evolution. The significant new contribution was identified as arising
from the backscattering of outgoing perturbations emerging near the
event horizon. The physical origin of such an outflux is nothing more
than the collapsing star which forms the black hole, and must surely
be present. The backscattered flux extends the instability of the CH
through the regime previously thought to be stable.  Furthermore our
analysis readily extends to the other proposed counter-examples to
strong cosmic censorship, such as accelerating black holes
\cite{gary}, or rotating black holes in de Sitter space \cite{chacha}.
Once again backscattering of initially outgoing modes provokes the
instability of the CH for otherwise stable configurations.
Thus one may conclude that there are no known
counter-examples to strong cosmic censorship within classical
general relativity coupled to reasonable matter.

\vskip 2ex
P.R.B. is supported by the Sherman Fairchild Foundation Inc. and NSF
Grant~AST-9417371.  R.C.M. is supported by NSERC of Canada, and by NSF
Grant~PHY94-07194. We would also like to acknowledge useful conversations
with D.M. Eardley and K. Thorne.

\end{document}